\documentclass[sigconf]{acmart}

\AtBeginDocument{%
  \providecommand\BibTeX{{%
    \normalfont B\kern-0.5em{\scshape i\kern-0.25em b}\kern-0.8em\TeX}}}

\setcopyright{none}
\acmConference[CHI '19 Workshops]{  ACM SIGCHI International Conference on Human Factors in Computing Systems}{May 2019}{Glasgow, UK}
\acmISBN{978-1-4503-9999-9/18/06}

\usepackage{subfigure}

\begin{document}

\title{Challenges of Designing HCI for Negative Emotions}

\author{Michal Luria}
\email{mluria@cs.cmu.edu}
\affiliation{%
  \institution{Human-Computer Interaction Institute}
  \city{Carnegie Mellon University}
}

\author{Amit Zoran}
\email{zoran@cs.huji.ac.il}
\affiliation{%
  \institution{School of Computer Science and Engineering}
  \city{Hebrew University}
}

\author{Jodi Forlizzi}
\email{forlizzi.cmu.edu}
\affiliation{%
  \institution{Human-Computer Interaction Institute}
  \city{Carnegie Mellon University}
}

\renewcommand{\shortauthors}{Luria et al.}
\begin{abstract}
Emotions that are perceived as ``negative'' are inherent in the human experience. Yet not much work in the field of HCI has looked into the role of these emotions in interaction with technology. As technology is becoming more social, personal and emotional by mediating our relationships and generating new social entities (such as conversational agents and robots), it is valuable to consider how it can support people's negative emotions and behaviors. Research in Psychology shows that interacting with negative emotions correctly can benefit well-being, yet the boundary between helpful and harmful is delicate. This workshop paper looks at the opportunities of designing for negative affect, and the challenge of ``causing no harm'' that arises in an attempt to do so. 
\end{abstract}



\maketitle

\section{Introduction \& Theoretical Background}
Negative emotions and behaviors are an inevitable aspect of people's lives. Although people tend to avoid them due to their unpleasantness, research in Psychology has recently shown that correctly engaging in negative expressions, such as anger, sadness or shame, can be important for our social relationships and for our own mental health~\cite{parrott2014positive}. 

As technology is becoming more personal and more social, it is critical to take into account how it might integrate negative emotions in interaction. Instead of attempting to immediately improve them or ignoring them altogether, interaction designers might consider embracing them in the interactions they design. 

Technology is often designed to support \textit{positive} emotions, yet it is not very common to encounter technology that helps people engage with emotions of sadness, anger or loneliness (as opposed to resolving them). This could be explained in part by the fact that happiness is frequently perceived as an important value to aspire to~\cite{mcmahon2006happiness}. In addition, people tend to feel \textit{aversion} towards negative emotions and believe that ``bad'' emotions are bad for you~\cite{henniger2014can,chentsova2014listening}.

Thus, it is not surprising that the design of technology reflects similar values and proposes to bring people together, make people happier, or help them be more productive and efficient. This is especially true for technology designed by industry. In addition to the natural human tendency to avoid negative emotions, they can undermine economic goals, which makes them even less desirable.

Aversion towards negative emotions is not ubiquitous, however. The philosophy of \textit{Wabi-Sabi} accepts that nothing is even perfect, permanent or complete, and sees the beauty in broken things~\cite{juniper2011wabi}. Buddhists engage in the destruction of \textit{sand mandalas} to emphasize the transience of life~\cite{bryant2003wheel}. More recently established rituals, such as the burning ritual at the culmination of the Burning Man festival, also illustrate the acceptance of destructive urges.

As technology gains a central role in shaping everyday life and is becoming increasingly social, perhaps there is a design space for interaction with social and personal negative emotions. However, due to its sensitivity, this space also introduces new challenges for HCI designers. 

\section{Designing Negative Emotion Interactions}
Designing for negative affect includes two significant challenges: (1) selecting a negative emotion and creating appropriate prototypes, and (2) engaging participants in interactions that are intended to surface this negative emotion. These challenges raise an ethical question---can designers and researchers explore and understand interactions in this design space while making sure they do not cause harm?

\begin{figure*}
  \centering
  \subfigure[Object 1]{
  \includegraphics[height=1.2in]{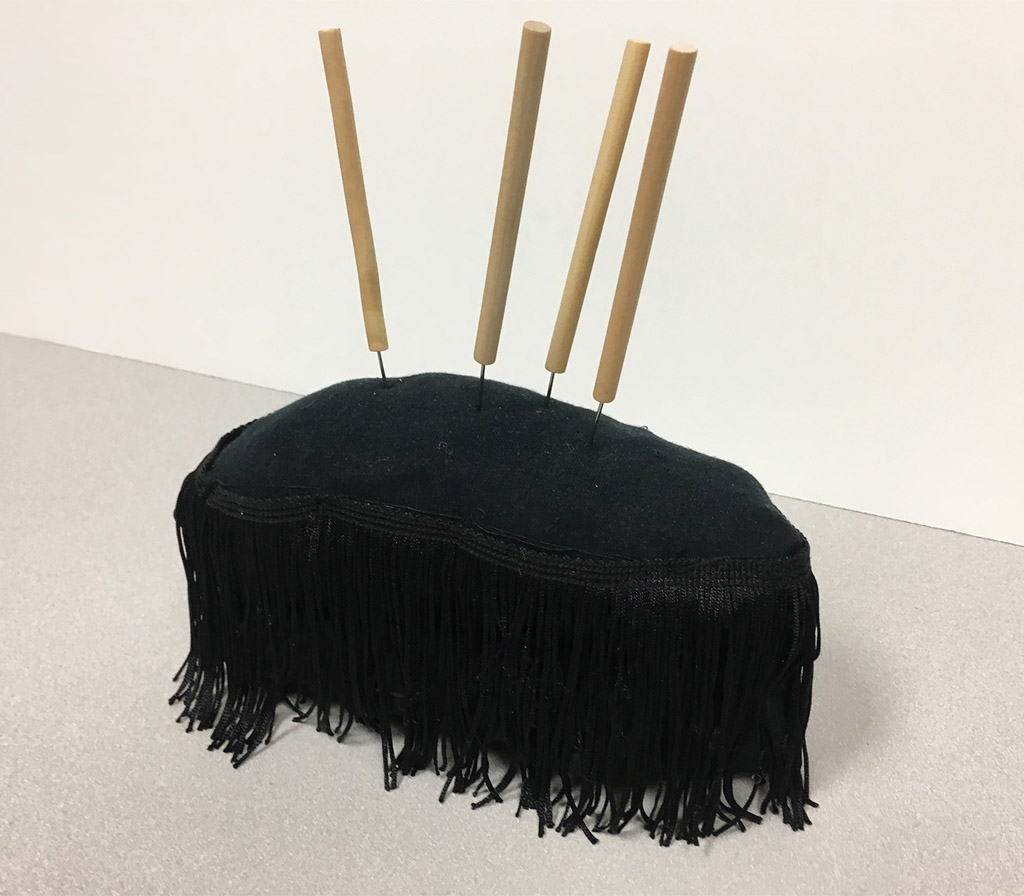}
  \label{fig:rob}}
  \qquad
  \subfigure[Object 2]{
  \includegraphics[height=1.2in]{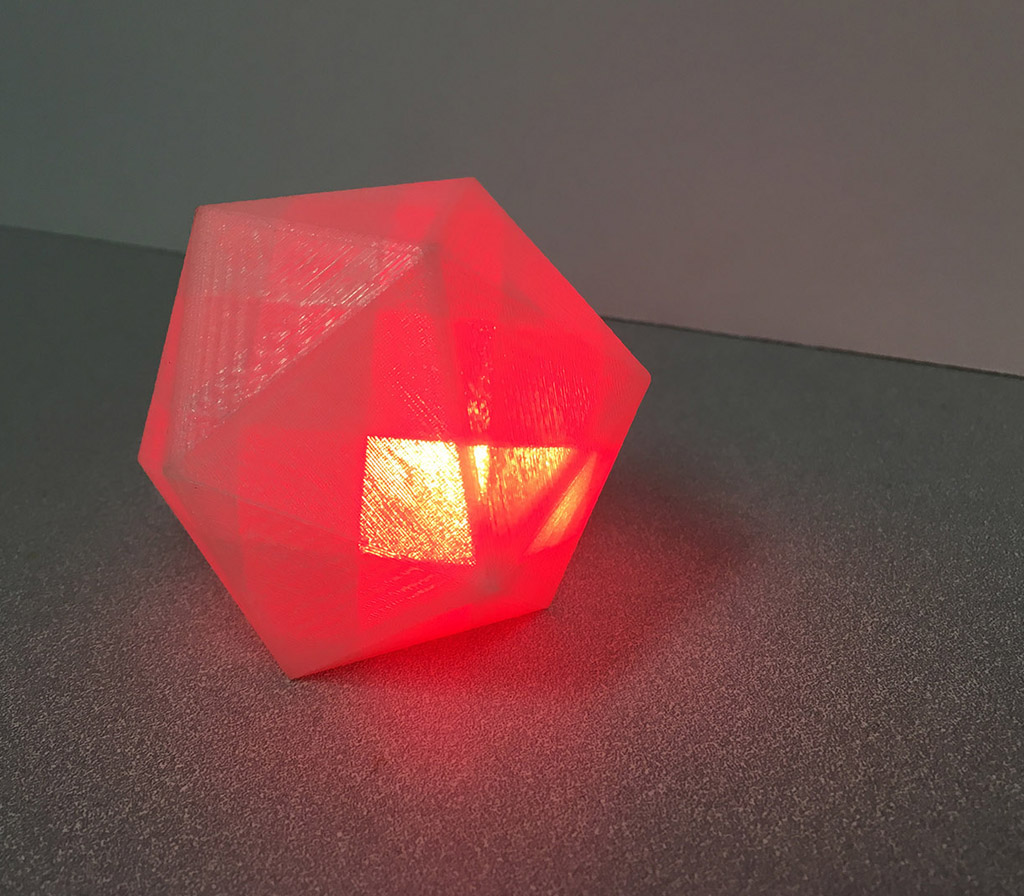}
  \label{fig:voc}}
  \qquad
  \subfigure[Object 3]{
  \includegraphics[height=1.2in]{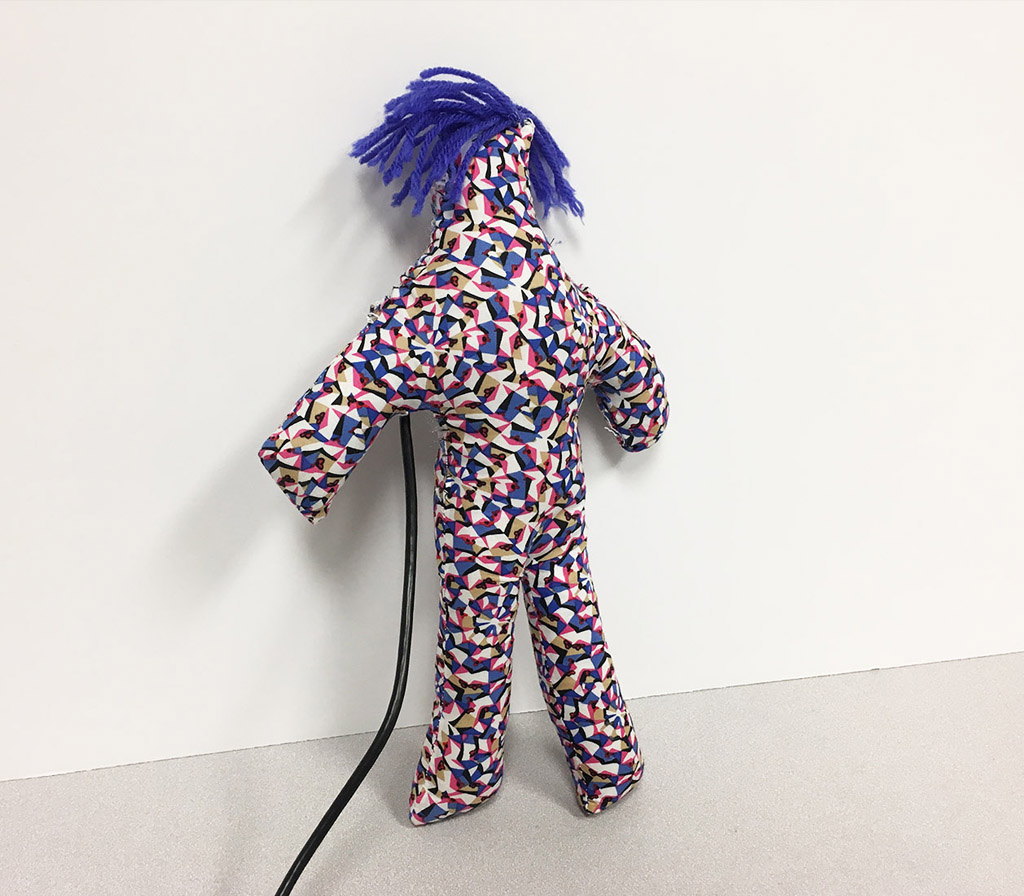}
  \label{fig:wal}}
  \qquad
  \subfigure[Object 4]{
  \includegraphics[height=1.2in]{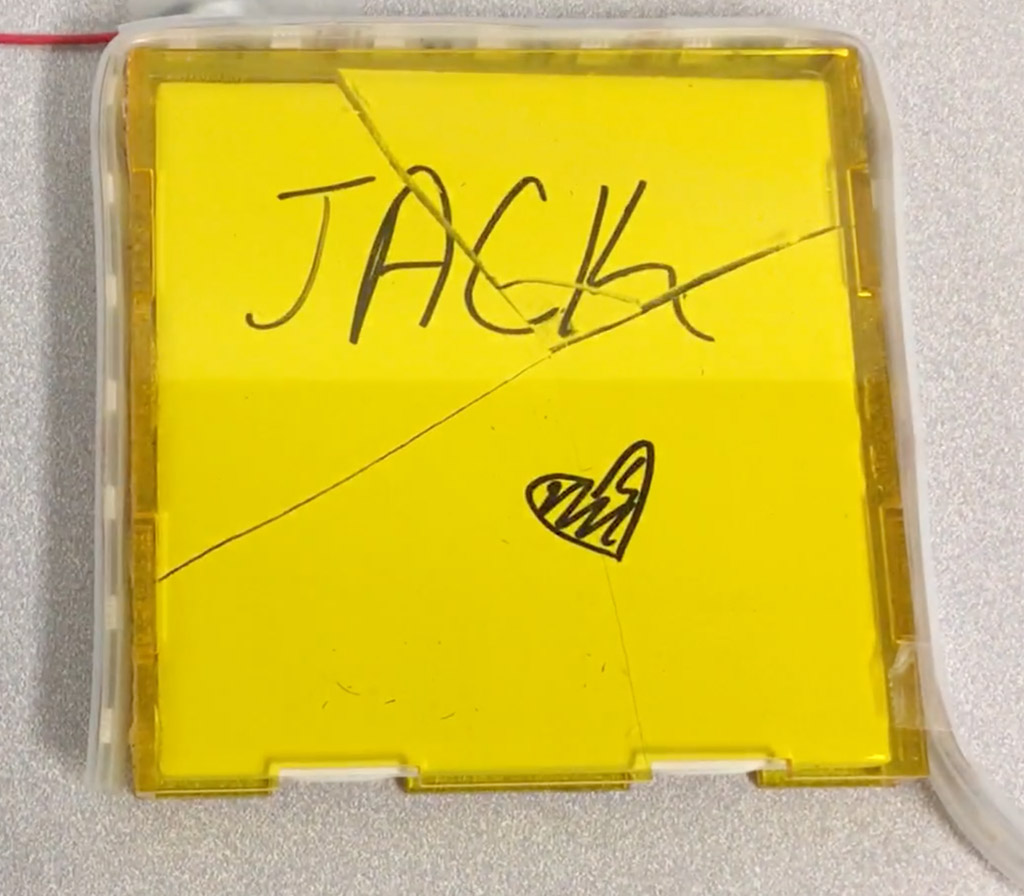}
  \label{fig:app}}
  \caption{\textit{Cathartic Objects}. Four objects that were designed for interactive expression of negative emotion through slow, reflective, forceful or verbal cathartic interactions.}~\label{fig:objects}
\end{figure*}

\subsection{Challenge 1: Designing for Negative Emotions}
The designer must consider several aspects of negative affect when choosing a specific topic for design. Negative expressions of emotion range from more destructive ones, such as anger, to ones that are simply unpleasant, like embarrassment. Each emotion has its own spectrum of severity. For instance, anger can be empowering, but can also be dangerous~\cite{hess2014anger}. The boundary between constructive and destructive designs can be very delicate, and is likely to vary from one person to another. Designers can build on research in psychology to design artifacts as responsibly as possible. Still, there is relatively little knowledge on the topic, and even less about how to design for it.

\subsection{Challenge 2: Evaluating Negative Emotions}
Once the prototype is complete, it is common for designers to evaluate the effectiveness of their design. Yet designing for negative affect implies negative user emotions will be involved, which might be enough to evoke the concern of academic institutional review boards. It is not very common for HCI researchers to promote negative emotions, and therefore it is unlikely to be approved for human-subject research. Even if approved, the researcher risks harming participants, or being denounced by the HCI research community for attempting to explore negative affect.

\section{Case Study: Cathartic Objects}
\textit{Cathartic Objects} explores the notion of interactive prototypes to support cathartic needs and behaviors. The original theory of catharsis argued that venting negative emotions can have a positive effect on one's mental state, as opposed to ``bottling it up inside''~\cite{breuer2009studies}. Although findings that confirm or refute this theory have been inconclusive, the idea of catharsis persists in people's beliefs in both historical and modern day expressions. Recent studies have also shown that venting can improve perceptions of fairness~\cite{liang2018righting} and can help relieve physical pain~\cite{stephens2009swearing}. Building on previous research and through an iterative design process, we designed four prototypes for catharsis that enable physical and vocal expression of negative affect (see Fig. 1).

\textit{Object 1} senses when it is poked with a sharp object. It responds in side-to-side gestures that signal it has absorbed the pain. When too many objects are inserted, it continues shaking until everything is removed, to encourage completion of a catharsis cycle.

\textit{Object 2} allows the user to verbally express frustration through cursing. The object recognizes cursing words, ``absorbs'' them, and ``re-purposes'' them as light energy.

\textit{Object 3} is a doll-like prototype that laughs in an irritating way when it senses the user is angry. Its goal is to encourage the user to physical express their emotional state using the doll. As the user hits the soft prototype against something, it stops laughing and re-evaluates the user's need for additional catharsis. 

\textit{Object 4} allows the user to create a personalized message, and then to destroy it. The user inscribes a ceramic tile and inserts it into the object to destroy with a hammer. As a result, the tile breaks and triggers a sequence of expressive light and sound, but is kept inside the object. The object allows the user to address a specific source of frustration without doing harm, and to use the artifact to document and reflect on their cathartic action.

\subsection{Challenge 1: Designing for Catharsis}
The first challenge was to research the topic of catharsis and to accordingly design artifacts that would be appropriate for that goal. We designed four \textit{Cathartic Objects} that allow a variety of emotional expressions---verbal, physical (forceful and gentle) and reflective---and that react in sound, light and movement. Although we drew on theoretical background, we also had to make design judgments about the implementation, as frequently necessary in a design process~\cite{gaver2012should}. For example, what seem to be the boundaries of interaction that allow catharsis to be beneficial, and not harmful. Or, what is the set of diverse prototypes that would support nuanced needs of venting. Due to the sensitivity of the topic, the iterative design process was not informed by pilots, but constrained to our own experiences. 

\subsection{Challenge 2: Evaluating Catharsis}
In the case of this project, it was not possible to simply transition to a standard evaluation process. Participants would be required to engage in negative emotions, and the benefits from doing so cannot be promised. Thus, it is unlikely that \textit{Cathartic Objects} can be evaluated in a human-subject study due to expected barriers in approval, recruitment, and ethical implications. The methodology planned for evaluation of this work is therefore an \textit{auto-ethnographical} study. One of the researchers will critically engage with the prototypes over an extended period of time, and would then analyze and report on their personal experiences~\cite{o2014gaining}. While this method may not be ideal, it might be the safest way to do so given the sensitivity of the research topic.

\section{Conclusion}
Negative emotions are a complex subject for most people, and as a result, a complex topic of research. Although psychologists show that negative affect is critical for well-being, the fact that people commonly believe it is best to avoid it~\cite{chentsova2014listening} makes it difficult to research in the field of HCI. By defining the challenges and applying them to the case-study of \textit{Cathartic Objects}, we learn that designers might be able to rely on literature and on their own judgment to sensibly design for negative emotions. However, evaluating the design still carries risks, and perhaps remains limited to auto-ethnographical research for the time being. 

\section{Acknowledgments}
This project was supported in part by funding from the Carnegie Mellon University Frank-Ratchye Fund For Art @ the Frontier.

\balance{} 

\bibliographystyle{ACM-Reference-Format}
\bibliography{sample}

\end{document}